\documentclass[final,5p,times,twocolumn]{elsarticle}

\usepackage{amssymb}

\usepackage{amsmath}

\usepackage{tabularx}
\usepackage{makecell}
\usepackage{multirow}

\usepackage[table]{xcolor} 

\usepackage{listings}
\usepackage{natbib}
\biboptions{authoryear,round,semicolon,sort&compress}
\bibpunct{(}{)}{;}{a}{,}{,}

\PassOptionsToPackage{hyphens}{url}
\usepackage{xurl}         
\usepackage[hidelinks]{hyperref}

\begin{document}
\begin{frontmatter}

\title{A Hybrid Machine Learning Framework for Improved Short-Term Peak-Flow Forecasting}

\author[1,2,4]{Gabriele Bertoli\corref{cor1}}
\ead{gabriele.bertoli@unifi.it}
\author[2]{Kai Schröter}
\author[3,4]{Rossella Arcucci}
\author[1]{Enrica Caporali}

\cortext[cor1]{Corresponding author}

\affiliation[1]{organization={Department of Civil and Environmental Engineering, University of Florence},
addressline={via di Santa Marta, 3},
city={Firenze},
postcode={50139},
country={Italy}}

\affiliation[2]{organization={Leichtweiß-Institute for Hydraulic Engineering and Water Resources, 
Division Hydrology and River Basin Management, Technische Universität Braunschweig},
addressline={Beethovenstr. 51a},
city={Braunschweig},
postcode={38106},
country={Germany}}

\affiliation[3]{organization={Department of Earth Science and Engineering, Imperial College London},
addressline={South Kensington Campus},
city={London},
postcode={SW7 2AZ},
country={United Kingdom}}

\affiliation[4]{organization={Data Science Institute, Imperial College London},
addressline={South Kensington Campus},
city={London},
postcode={SW7 2AZ},
country={United Kingdom}}

\begin{abstract}

Reliable river flow forecasting is an essential component of flood risk management and early warning systems. It enables improved emergency response coordination and is critical for protecting infrastructure, communities, and ecosystems from extreme hydrological events. Process-based hydrological models and purely data-driven approaches often underperform during extreme events, particularly in forecasting peak flows. To address this limitation, this study introduces a hybrid forecasting framework that couples Extreme Gradient Boosting (XGBoost) and Random Forest (RF). XGBoost is employed for continuous streamflow forecasting, while RF is specifically trained for peak-flow prediction, and the two outputs are combined into an enhanced forecast. The approach is implemented across 857 catchments of the LamaH-CE dataset, using rainfall and discharge observations at 6-hour resolution. Results demonstrate consistently high skill, with 71\% of catchments achieving a Kling–Gupta Efficiency (KGE) greater than 0.90. Peak-flow detection reaches 87\%, with a false-alarm rate of 13\%. Compared to the European Flood Awareness System (EFAS), the framework achieves lower peak-magnitude errors, fewer false alarms, and improved streamflow and peak-flow forecasting accuracy. The proposed framework is computationally lightweight, scalable, and easily transferable across watersheds, with training times of only seconds on standard CPUs. These findings highlight the potential of integrating hydrological understanding with efficient machine learning to improve the accuracy and reliability of operational flood forecasting, and outline future directions for hybrid hydrological–machine learning model development.

\end{abstract}


\end{frontmatter}

\newpage
\section{Introduction} \label{section:introduction}

Reliable river flow forecasting is essential for effective civil protection, flood control, water resources management, and ecosystems preservation. Within this context, peak flow forecasting is especially critical, as it underpins emergency management by providing the timely information required to anticipate and mitigate flood hazard exposure and vulnerability of communities, assets, infrastructure, and natural systems. Reliable peak flow forecasting is therefore essential for enabling preventive measures, optimizing emergency response strategies, and ultimately minimizing human and economic losses during extreme hydrological events. Process-based hydrological models constitute a fundamental part of flood forecasting systems, describing the complex physical processes behind the rainfall-runoff transformation. In parallel, artificial intelligence methods such as machine learning, deep learning, and data assimilation have recently emerged as powerful alternatives or complements e.g.~\citep{Gao2020}. Bridging the two worlds, there are new hybrid frameworks that integrate hydrological principles with artificial intelligence, as in~\citep{Nearing2024, Xu2024}. 
However, flood forecasting techniques developed over recent decades show both strengths and limitations, as outlined by~\citep{Mishra2022, Das2022, Nearing2021}. Their performances are often affected by the large number of uncertainties, as pointed out by~\citep{Merz2021, MKemter2020, MIBrunner2020}, that can arise from input data, watershed characteristics or be inherent to extreme and rare events~\citep{Han2025, BFang2024, AMontanari2024, VDNguyen2024, Merz2024}. Moreover, peak flow forecasts remain particularly challenging due to the intrinsic complexity of hydrological processes and the scarcity of recorded historical data~\citep{Brunner2021, Kreibich2022}, despite being the most important target of a flood forecasting system. \\ 

The goal of this study is to develop a fast, efficient, and accurate machine learning flood forecasting framework, achieving the most accurate and reliable forecasts, while maintaining computational efficiency and model simplicity. The objective is to deliver reliable short-term river flow forecasts, with a particular emphasis on the accurate prediction of peak flow events, by integrating domain-specific hydrological understanding into the conceptualization, data selection, and adaptation of machine learning algorithms. 
In this study we propose a novel flood forecasting framework based on the coupling of Extreme Gradient Boosting (XGBoost) and Random Forest (RF) machine learning methods, testing it on the LamaH-CE large sample hydrology dataset~\citep{essd-13-4529-2021}.

To the best of our knowledge, this is the first instance in which this combined machine learning framework has been conceptualized and implemented, and the first time that Random Forest has been applied to sub-daily river discharge forecasting.

\subsection{Paper Structure}

The rest of this paper is organized as follows. Section~\ref{section:related works} provides a brief contextualization and a concise review of existing literature on forecasting technologies. Section~\ref{section:data} outlines the datasets employed in the research, while data preprocessing is detailed in subsection~\ref{subsection:preprocessing}. The proposed methodology is described in Section~\ref{section:method}, with subsection~\ref{subsection:Model} presenting the overarching framework. Subsequently, subsections~\ref{subsection:XGBoost} and~\ref{subsubsection:lag-roll} elaborate on the implementation of XGBoost and the modelling of temporal dependencies, respectively. Subsection~\ref{subsection:RF} details the Random Forest configuration, and subsection~\ref{subsection:integration} describes the integration of the two methods within the forecasting framework, with operational implementation specifics provided in subsection~\ref{subsubsection:operational}. Section~\ref{section:metrics} outlines all evaluation metrics used to assess the forecasting performance, and in subsection~\ref{subsection:method-power-estimate}, we described the required energy estimation method. In Section~\ref{section:results} we present the findings objectively, quantified through the selected metrics. Finally, Section~\ref{section:discussion} interprets the results from both hydrological and machine learning perspectives. The paper concludes with Section~\ref{section:conclusion}. 

\section{Related works and broader context} \label{section:related works}
In this section, a brief review of the current limitations and solutions provided by the literature is presented.\\
Recent advancements in machine learning and hybrid frameworks have addressed these challenges as evidenced by the diverse methodologies and innovations highlighted in the following studies.
For example,~\citep{Gao2020} demonstrate the effectiveness of deep learning architectures such as GRU and LSTM for rainfall-runoff prediction, while~\citep{Ding2020} introduce an interpretable spatio-temporal attention LSTM tailored to flood forecasting. In addition,~\citep{Nevo2022} present the deployment of machine learning models in an operational flood forecasting framework, and~\citep{Slater2023} explore hybrid approaches that combine climate predictions with AI models to improve large-scale hydrological forecasts. Reviews such as~\citep{Cheng20231361} synthesize progress at the intersection of machine learning, data assimilation, and uncertainty quantification in dynamical systems, while frameworks developed by~\citep{Arcucci20211} and \citep{lever2025facing} highlight how deep data assimilation can embed domain knowledge into learning algorithms to enhance robustness.
Hybrid frameworks which merge data-driven methodologies with process-based understanding have demonstrated significant potential to enhance forecasting accuracy and address the shortcomings of purely data-driven or purely process-driven models. For instance,~\citep{Nearing2024} introduce physics-guided machine learning frameworks that improve continental-scale flood predictions, while~\citep{Tripathy2024} explore multi-source data fusion to strengthen spatio-temporal generalization.~\citep{Xu2024} propose a hybrid model that explicitly couples process-driven and data-driven components for improved real-time forecasting, and~\citep{Yang2020} demonstrate the value of hybridization for short-term flood prediction. Beyond methodological advances,~\citep{Bennett2023} emphasize the operational benefits of such frameworks in decision support, whereas~\citep{Barzegar2021} highlight their ability to handle non stationarity in hydrological processes. Finally,~\citep{Wang-Bertoli-ICCS25, 4d-latent} showcase scalable hybrid data assimilation techniques that integrate convolutional autoencoders with LSTM architectures, extending applicability to computationally demanding forecasting scenarios. \\
However, machine learning models for hydrological forecasting generally benefit from dynamic inputs (e.g., rainfall time series), as these explicitly represent the temporal variability of hydrological processes. In contrast, static layers that describe watershed characteristics (e.g., geology, elevation) lack any temporal dimension, and their repeated input across time steps may mask or bias the learning of temporal dependencies. Machine learning-based methods also require large volumes of high-quality data, which may not always be readily available or consistently maintained. Moreover, the computational demand required to train and execute such models can be significant, posing practical constraints. Finally, fine-tuning of model parameters remains critical to achieve accurate forecasts, yet this step is often resource-intensive and complex. 
In addition, extreme events such as flood peaks are rare, and usually, even in long-term streamflow records, only a limited number are observed. This scarcity presents a significant challenge for both process-based and machine learning-based forecasting methods, which are trained on large datasets dominated by normal or moderate flow conditions, but with only a small subset of extreme flood events. The learning is then unbalanced~\citep{Salaeh2026, Moniz2017} with respect to the streamflow, and the forecasting abilities for extreme events are necessarily always lower than for smaller floods and usual river flow. To address this imbalance and data scarcity, recent studies have proposed promising methods, including resampling-driven machine learning models~\citep{Salaeh2026}, deep generative modelling~\citep{Sattari2025}, hybrid neural networks such as SCS-LSTM~\citep{Ximin2025}, deep spatio-temporal models like R2RNet~\citep{Dhankhar2025}, attention-enhanced deep learning~\citep{Chen2025}, and surrogate modelling frameworks integrating net rainfall calculations~\citep{Farfan2025}.

\section{Data Acquisition and Preprocessing} \label{section:data}

Reliable, detailed, and extensive spatio-temporal datasets are fundamental for analysing hydrological processes such as river flow generation. In this work, we adopted a parsimonious modelling setup, leveraging only rainfall and river discharge observations, to evaluate the feasibility of providing accurate flood forecasts under minimal data requirements.

\subsection{River Discharge Data} \label{subsection:river data}

\begin{figure*}
 \centering
 \includegraphics[width=0.75\linewidth]{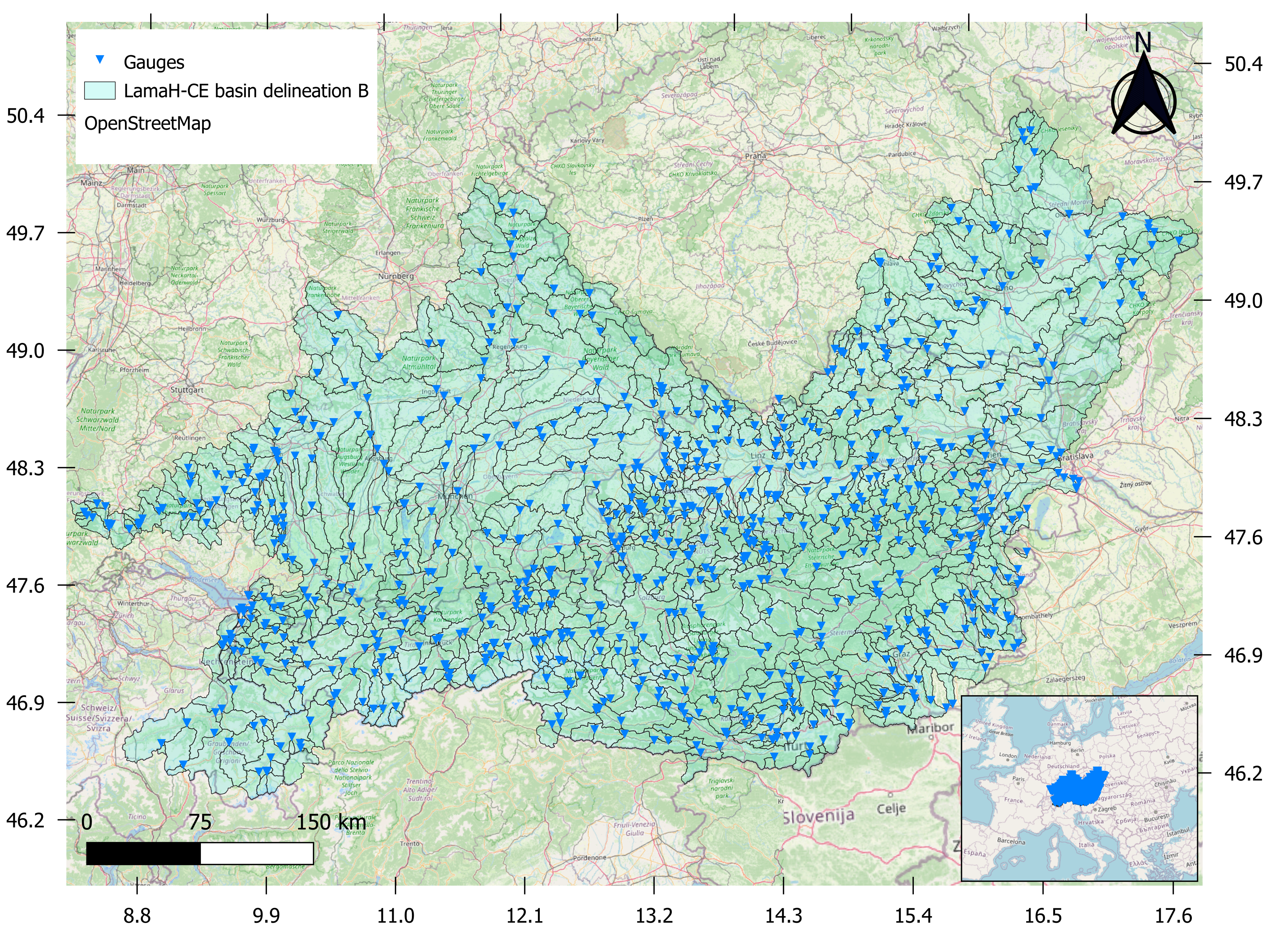}
 \caption{Location of the 859 LamaH\textendash CE gauged catchments (following the LamaH-CE's delineation B, light blue polygons) and their gauges (blue triangles). The spatial domain spans approximately $8.8^{\circ}$-$17.6^{\circ}$E and $46.2^{\circ}$-$50.4^{\circ}$N, covering about \(170{,}000~\text{km}^2\) mainly across Germany, Switzerland, Austria, and the Czech Republic. The lower right box shows the location of the domain within Central Europe. Background: OpenStreetMap.}
 \label{fig:Lamah_dataset}
\end{figure*}

To select a suitable database for this study, we evaluated several publicly available river discharge datasets. Given our objective of modelling flood peaks, a sub-daily temporal resolution was essential to accurately represent such events. To enhance the potential for future applications of our study and to effectively capture watershed variability, only large-scale spatial datasets have been analysed, despite their limited availability. We adopted the LamaH-CE dataset~\citep{essd-13-4529-2021}, due to its scalable, comprehensive, detailed, and controlled collection of high-quality hydrological observations and morphological data. LamaH-CE stores river discharge observations for 859 catchments at hourly resolution, (Figure ~\ref{fig:Lamah_dataset}) over approximately 170000 km\textsuperscript{2}, mainly across Germany, Switzerland, Austria and Czech Republic, with a time range that spans from 1981 to 2017.

As a benchmark, we evaluated the European Flood Awareness System (EFAS)~\citep{EFASv5_2023}, which is the official early warning system for floods in Europe. It represents the state of the art in flood forecasting at a continental scale in Europe providing spatially continuous river discharge information. It has a spatial resolution of 1 arcmin and a temporal resolution of 6 hours. This system is based on the LISFLOOD~\citep{lisflood} model and excels in long-term operational forecasting for large, cross-boundary watersheds. It is less suitable for medium to small watershed assessments. Nevertheless, it also covers ungauged basins, offering useful information even where data are scarce.

\subsection{Rainfall Data} \label{subsection:rainfall}

Similarly to the selection procedure carried out for the river discharge dataset, finding historical rainfall data for this study involved accurate analysis of the characteristics of publicly available datasets to find the best compromise between: (1) spatial resolution, (2) temporal resolution, (3) time coverage, and (4) data reliability. The objective was to maximize all these characteristics. We considered ground observations like gauges and radars more reliable than reanalysis products, which have been demonstrated to be less suitable for extreme rainfall events~\citep{Zhu2017,Rhodes2015,Odon2019,Guo2022,Hu2020,Wang2023}. These events, which significantly influence river flow modelling, were indeed prioritized in the analysis. We selected the EMO-1 dataset~\citep{EMO2025} from the Copernicus Emergency Management Service for Europe. EMO-1 is a gridded dataset with 1 arcminute spatial resolution, critical for capturing rainfall differences in catchments, especially in small to medium-scale basins. It covers the period from 1990 to 2022 and stores sub-daily data at a 6-hour time resolution, providing detailed information on the temporal evolution of rainfall, which is particularly relevant for medium to small-scale watersheds with short times of concentration. It combines multiple data sources, including satellite observations, radar, and in-situ rain gauges. The raw observations have undergone a rigorous quality control process, followed by SPHEREMAP spatial interpolation~\citep{Willmott1985, Shepard1968} and Yamamoto kriging~\citep{Yamamoto2005}, to estimate both the rainfall values and associated uncertainty for each grid cell, enhancing the overall reliability as input for river flow modelling~\citep{Huang2019, Gupta1999}.

\subsection{Data preprocessing} \label{subsection:preprocessing}

For the comparison of EFAS forecasts at LISFLOOD model grid cells resolution and LamaH-CE river gauges, an automated approach was adopted to link both data sets, as introduced in~\citep{Wang-Bertoli-ICCS25}. The method, based on the hydrological concept of watershed, selected the EFAS cell with the highest streamflow within each watershed (i.e. the best approximation possible to the streamflow measured by the river gauge at the watershed's outlet), ensuring accurate alignment of the two datasets despite discrepancies in DEM resolution and river network geometry. However, some peculiar river or morphological characteristics may have caused misalignments, which we reduced through a manual check. Since rainfall data from EMO-1 has 6 hour time resolution, and EFAS river discharge values are made available as the average over the last six hours, we averaged the LamaH-CE river flow observation values to match these intervals, ensuring consistency and minimizing additional uncertainty. The watersheds delineated in the LamaH-CE dataset (delineation B) were used as reference units, and the rainfall data were clipped over each catchment. We then spatially aggregated rainfall values (mm/6-hour) within each watershed domain to obtain a single value of areal rainfall for each time step. 

\section{Methods} \label{section:method}  

\begin{figure}[h!]
 \centering
 \includegraphics[width=0.8\linewidth]{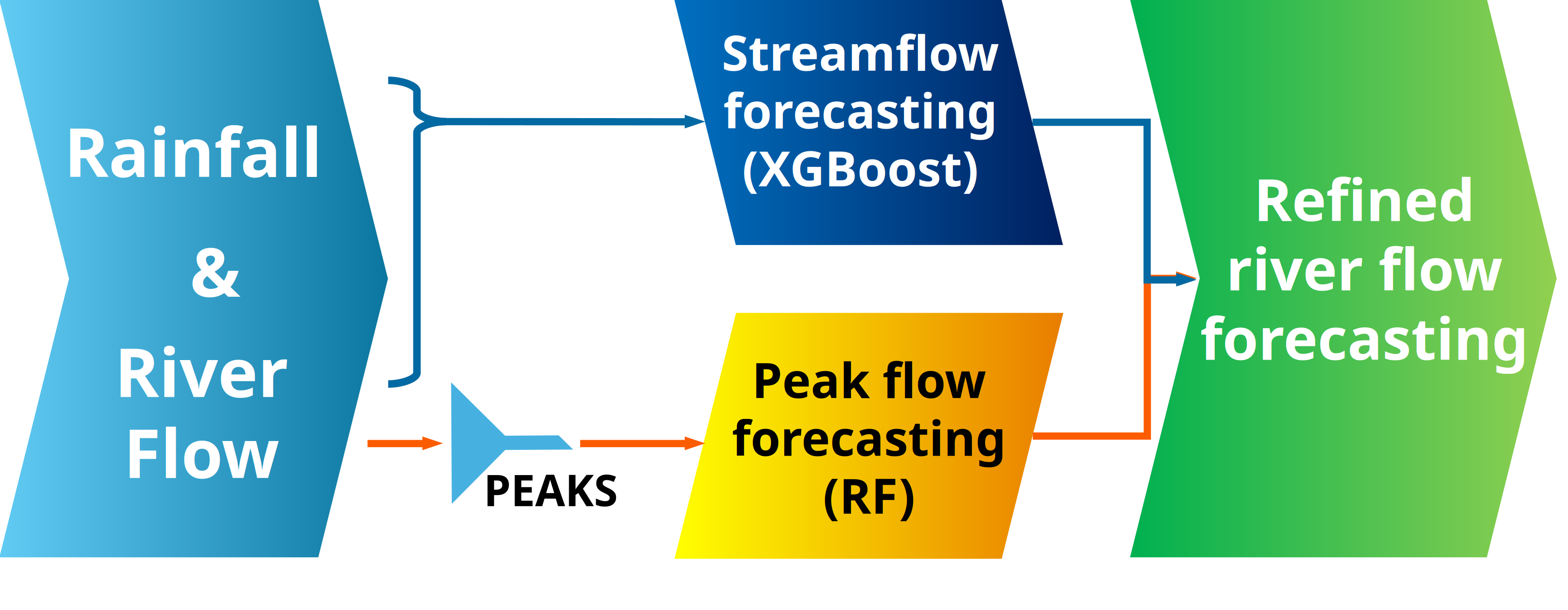}
 \caption{Flowchart of combining XGBoost streamflow and RF peak flow forecasting using rainfall and river flow observations to provide a refined river flow forecasting framework}
 \label{fig:flowchart}
\end{figure}

This section describes the methods we implemented in this research (see Figure~\ref{fig:flowchart}), such as XGBoost for river flow forecasting and Random Forest for peaks forecasting, and their integration in the forecasting framework.

\subsection{XGBoost-RF river flow forecasting framework} \label{subsection:Model}  

We combined the XGBoost and RF machine learning methods for accurately forecasting both streamflow and peak flows. 
Specifically, the XGBoost model focused on forecasting streamflow, while the RF model was dedicated to peak flows forecasting. The results from the two models are subsequently merged to obtain refined forecasts, as illustrated in Figure~\ref{fig:flowchart}. We applied the developed forecasting method to 857 watersheds in the LamaH-CE domain. In the following sections the development and implementation processes are described.

\subsection{Extreme Gradient Boosting (XGBoost)} \label{subsection:XGBoost}
XGBoost is a scalable machine learning library based on the gradient boosting structure, introduced by Tianqi Chen ~\citep{Chen_2016} and subsequently refined in the context of the Distributed (Deep) Machine Learning Community (DMLC), as reflected in its official project documentation~\citep{DMLC_Website} and contributor list~\citep{XGBoost_Contributors}. It is a decision-tree-based ensemble learning algorithm that builds models in a sequential, additive way to minimize forecast error: at each iteration, a new decision tree is trained to forecast the residuals (errors) of the existing ensemble. XGBoost has a better optimization and algorithm setup with respect to traditional gradient boosting techniques, and it uses second-order gradients for a better convergence and forecasting accuracy. It also implements advanced refinements to control overfitting and handling missing values. Computational performance is improved also (but not only) thanks to the parallelized tree construction capability at the feature level, which enables XGBoost to build trees more efficiently by distributing the computation of split points across multiple CPU cores during training.
XGBoost is not specifically designed for sequential operations (e.g., time series forecasts), but re-framing the problem as a supervised learning task effectively enabled its application to this task.

\subsubsection{Implementation of temporal dependencies in XGBoost: lag features and rolling statistics} \label{subsubsection:lag-roll}

As introduced above, XGBoost treats each input feature as an independent variable, lacking an understanding of temporal sequence. Therefore, it was necessary to explicitly encode time-related information into the input space through appropriate feature engineering techniques. Two widely used procedures for this objective are the construction of lag features and the computation of rolling statistics~\citep{Cerqueira2021, Widodo2016}. These methods effectively embed temporal structure into the dataset, allowing the model to learn from historical dynamics and temporal patterns. 
We implemented these methods due to the specific characteristics of the forecasting task and the available data. For the intended one-step ahead flood forecasting lag features provide information on recent system states. Rolling statistics describe short-term variability and trends - both of which are relevant for representing the dynamics associated with the formation of peak flows and extreme events. These techniques allow for an effective representation of short-term hydrological processes, without adding complexity, and thus help preserve the interpretability of the model. 
In detail, lag features provide discrete access to prior states of the system, thus capturing autocorrelations, delayed effects, and other dynamic dependencies between past and present observations. This gives the model explicit short-term memory without compressing information across time. Formally, given a time series $x_t$, a lag feature of order $l$ is defined as $x_{t-l}$. In this study, lagged versions of both rainfall and river discharge were generated. The number of relevant lags depends on the physical characteristics that control how quickly a watershed responds to rainfall. Given the large number of watersheds analysed, we used a fixed lag of 4 time steps for the rainfall, and 3 for the discharge, which were calibrated on a subset of 3 catchments (carefully selected to well represent hydrological diversity) and then applied uniformly across all watersheds in the study area. 
Rolling statistics instead describe the behaviour of a variable over a moving window of length $w$. Unlike lag features, which supply single past observations, rolling statistics condense multiple past points into a single descriptor, capturing short-term trends, persistence, and variability. These statistics are less sensitive to noise and more effective at capturing local patterns over time. In this study, rolling means and standard deviations were calculated for both short and intermediate windows. After calibration as described above, a rolling mean window of 4 steps is set for the discharge, such that at each time step $t$, the computed statistic (e.g., mean or standard deviation) summarizes the values from time steps $t-3$ to $t$, included. A window of 5 steps is defined for the standard deviation. For the rainfall we calibrated a 2 time steps rolling mean window and a 5 time steps window for the standard deviation. After computing lagged values and rolling statistics, they were passed to XGBoost as standard tabular inputs.

\subsubsection{XGBoost hyperparameters tuning and custom performance metric setup}

We tuned the XGBoost regressor hyperparameters to achieve an optimal balance between model complexity and predictive accuracy, setting a learning rate of 0.1 with a maximum tree depth of 5 levels, subsampling at 95\% for each iteration, feature subsampling at 90\% at each split, a minimum child weight of one, no L1 regularization, moderate L2 regularization, and no minimum loss reduction requirement for further partitioning on leaf nodes. These settings collectively ensure the model captures non-linear relationships while maintaining robustness and generalization capability. In detail, the lower learning rate allows for more iterations to converge without overfitting, while limiting tree depth controls complexity and helps capture non-linear patterns. Subsampling and feature subsampling introduce randomness and decorrelation among trees, reducing overfitting and improving generalization. Setting a minimum child weight regularizes the model by preventing overly small leaf nodes that might fit noise. L1 and L2 regularization penalize large coefficients, simplifying the model and enhancing generalization. Allowing more complex trees without additional loss reduction requirements better captures intricate patterns in the data.
A custom performance metric - the Kling-Gupta Efficiency (KGE) - is implemented to guide model optimization, as it better reflects hydrological model accuracy than other metrics such as NSE, RMSE or \(R^2\). Details for these metrics are provided in the dedicated section~\ref{section:metrics}.

\subsection{Random Forest (RF) for Peak forecasting} \label{subsection:RF}

Random Forest also belongs to the family of ensemble learning algorithms based on decision trees, as XGBoost, but it follows a distinct approach. It implements a technique known as bagging, where multiple trees are built independently and in parallel on bootstrapped subsets of the data, and their outputs are then aggregated to reduce variance. RF also introduces another layer of randomness, considering a random subset of features at each split, which, in addition to the bootstrapped sampling, helps reduce overfitting. It is a non-parametric model capable of capturing complex and strongly non-linear relationships between the input features and the target variable, even with noisy or smaller subsets of data. 
The peak flows were modelled using a dedicated Random Forest regressor, with a training dataset comprising only peaks as described in the following subsection~\ref{subsubsection:peak_def}. This selective approach ensures that the RF model explicitly learns flow behaviour and rainfall patterns during and before extreme events, rather than general flow variability. 

\subsubsection{Peak flow definition} \label{subsubsection:peak_def}

A peak discharge event was defined as any observation exceeding the empirical 99.9\textsuperscript{th} percentile of the river discharge distribution. We selected the 99.9\textsuperscript{th} percentile to correspond with rare and extreme high-flow conditions. This choice is consistent with standard practices in flood risk assessment and early warning systems, which often concentrate analyses on the upper tail of flow distributions. For each time step, we assigned a binary flag indicating whether the discharge exceeded this threshold. Additionally, we defined a semi-continuous variable, \texttt{peak\_magnitude}, representing the severity of each peak event as the extent by which it surpassed the 99.9\textsuperscript{th} percentile, with positive values signifying peak intensities and zeros indicating non-peak periods. Formally, this is expressed as:

\begin{equation}
\texttt{peak\_magnitude}_t =
\begin{cases}
Q6h_t - \text{threshold}, & \text{if } Q6h_t > \text{threshold}, \\
0, & \text{otherwise}.
\end{cases}
\end{equation}

where \( Q6h_t \) denotes the average of the river discharge observations of the last six hours with respect to time step \( t \), as described in section~\ref{subsection:preprocessing}. 

\subsubsection{Random Forest hyperparameters tuning and implementation} \label{subsubsection:RF implementation}

We tuned the RF regressor hyperparameters for an optimal balance between model complexity and predictive accuracy: 700 decision trees with a maximum depth of eight levels, a minimum of five samples per leaf node, and feature subsampling at 80\% at each split. The choice of 700 estimators provides sufficient ensemble diversity to capture non-linear relationships while mitigating the marginal gains observed with larger ensembles. Limiting tree depth to eight levels constrains model complexity, thus reducing overfitting risk while preserving essential hierarchical interactions. A minimum of 5 samples per terminal node decreases variance by preventing splits on small data subsets. Feature subsampling at 80\% introduces additional decorrelation among trees, reducing susceptibility to predictor dominance and overfitting risk (since the features are randomly selected in the 80\% subset). 

\subsection{XGBoost and RF forecasts integration} \label{subsection:integration}

To obtain the final forecast, the XGBoost model's forecasts are used as a baseline for streamflows. Identified peak indices within the test set are then updated by incorporating the RF peak model outputs, denoted \(\hat{y}^{\text{peak}}\). Specifically, for each such peak of index \(i\), the forecast is modified as follows:
\[
\hat{y}_{i} = \hat{y}^{\text{main}}_{i} + 0.95 \times \hat{y}^{\text{peak}}_{j}
\]
where \(j\) corresponds to the matching peak forecast. The 0.95 value was obtained after calibration on a subset of catchments and extended to all the watersheds. The obtained results confirmed that this is a useful estimate, but site-specific calibration is possible if further tuning is required. For all other indices, the forecasts remain unchanged:
\[
\hat{y}_{i} = \hat{y}^{\text{main}}_{i}
\]

This combined strategy ensures the model captures both general hydrological behaviour and peak deviations.

\subsubsection{Operational implementation of XGBoost and RF model for flow forecasting: summary} \label{subsubsection:operational}

A reproducible Python workflow was prototyped and iteratively refined on a randomly selected subset of hydrologically diverse LamaH-CE catchments. This pilot phase served to (1) stabilize the feature set selection (lags and rolling windows), (2) fix XGBoost and Random Forest hyperparameters for balanced performance across different watersheds, and (3) define the integration strategy between streamflow and peak values. Once this modelling chain structure was established, it was applied unchanged to 857 selected watersheds. Each watershed-specific model was trained on its own discharge-rainfall time series, ensuring local calibration while keeping the computational load low thanks to the lightweight structure and the fully automated pipeline.
Operationally, the implementation followed these steps:
\begin{enumerate}
 \item \textbf{Pilot tuning:} select a random set of basins; develop and test the full pipeline; choose a single, robust configuration of features, hyperparameters, and combination weights.
 \item \textbf{Structure setup:} lock the modelling structure (feature engineering and hyperparameters) to avoid ad hoc tuning for individual catchments.
 \item \textbf{Batch deployment:} load basin-specific data, train the two models, generate refined flood forecasts using the locked setup from step 2, and loop over all basins.
 \item \textbf{Quality control and storage:} compute evaluation metrics, save forecasts, and metrics.
\end{enumerate}

\section{Evaluation metrics} \label{section:metrics}

In this section, we list and introduce the metrics that have been adopted to evaluate multiple aspects of the performances of the XGBoost - RF forecasting framework. 

\subsection{Nash-Sutcliffe Efficiency (NSE)}

The Nash-Sutcliffe Efficiency (NSE)~\citep{Nash1970} is a widely used performance metric in hydrological modelling for evaluating the skill of discharge simulations. It measures how well the simulated time series matches the observed data in terms of magnitude and temporal dynamics.
The NSE is defined as:
\begin{equation}
NSE = 1 - \frac{\sum_{t=1}^{n} (Q_{\text{sim}, t} - Q_{\text{obs}, t})^2}{\sum_{t=1}^{n} (Q_{\text{obs}, t} - \overline{Q}_{\text{obs}})^2}
\end{equation}
where $Q_{\text{sim}, t}$ and $Q_{\text{obs}, t}$ are the simulated and observed discharges at time $t$, and $\overline{Q}_{\text{obs}}$ is the mean of the observed discharges.
A NSE value of 1 indicates a perfect match between simulations and observations, while a value of 0 indicates that the model is performing as good as using the mean of the observations. Negative values indicate that the model performs worse than using the mean of the observations.

\subsection{Kling-Gupta Efficiency (KGE) and Modified KGE} \label{subsection:kge}

The Kling-Gupta Efficiency (KGE)~\citep{gupta2009decomposition} is a performance metric developed to overcome some limitations of the Nash-Sutcliffe Efficiency by decomposing model performance into correlation, bias, and variability components. It allows a more balanced assessment of hydrological model behaviour, particularly when evaluating flow magnitudes.
The KGE is defined as:
\begin{equation}
KGE = 1 - \sqrt{(r - 1)^2 + (\beta - 1)^2 + (\gamma - 1)^2}
\end{equation}
where:
\begin{itemize}
 \item $r$ is the linear correlation coefficient between simulated and observed flows,
 \item $\beta = \frac{\mu_s}{\mu_o}$ is the bias ratio, with $\mu_s$ and $\mu_o$ the mean simulated and observed flows, respectively,
 \item $\gamma = \frac{CV_s}{CV_o}$ is the variability ratio, where $CV$ denotes the coefficient of variation (standard deviation divided by the mean).
\end{itemize} 
The KGE ranges from $-\infty$ to 1, with a value of 1 indicating a perfect match between the simulated and observed data in terms of correlation, bias, and variability.

Also, the modified version of KGE, often referred to as KGE', is considered in this study since it's one of the metrics adopted in the official EFAS documentation. It adjusts the variability component to use the standard deviation ratio instead of the coefficient of variation, enhancing sensitivity to differences in variability when the mean of the simulated flow significantly differs from the mean of the observed flow~\citep{Kling2012}. It is defined as:
\begin{equation}
KGE' = 1 - \sqrt{(r - 1)^2 + (\beta - 1)^2 + (\gamma' - 1)^2}
\end{equation}
where:
\begin{equation}
\gamma' = \frac{\sigma_s}{\sigma_o}
\end{equation}
with $\sigma_s$ and $\sigma_o$ being the standard deviation of simulated and observed flows, respectively.

\subsection{Relative Peak Error (RPE)}

The Relative Peak Error (RPE) was selected to evaluate the model's ability to forecast the magnitude of peak discharges during flood events, in a scale-independent way. It is defined as:
\begin{equation}
RPE = \frac{Q_{\text{sim, peak}} - Q_{\text{obs, peak}}}{Q_{\text{obs, peak}}}
\end{equation}
where $Q_{\text{sim, peak}}$ and $Q_{\text{obs, peak}}$ represent the simulated and observed peak discharges, respectively.
Peaks have been defined coherently with the definition adopted in the model, described in section~\ref{subsubsection:peak_def}.
A positive RPE value shows that the model overestimates the peak, while a negative value indicates underestimation. This is useful for identifying systematic biases in the forecasting system.
The absolute RPE is calculated by taking the absolute value of the RPE:
\begin{equation}
|RPE| = \left| \frac{Q_{\text{sim, peak}} - Q_{\text{obs, peak}}}{Q_{\text{obs, peak}}} \right|
\end{equation}
This version focuses on the magnitude of the forecast error regardless of its sign and is useful for assessing the overall accuracy of peak forecasts across multiple events.

\subsection{Peak Timing Error}

We computed the Peak Timing Error metric to evaluate the model’s accuracy in forecasting the timing of flood peaks. This metric specifically quantifies the temporal deviation between observed and forecasted peaks and it's defined as follows:

\begin{equation}
\text{Peak Timing Error} = T_{\text{sim, peak}} - T_{\text{observations, peak}}
\end{equation}

where:
\begin{itemize}
 \item $T_{\text{model, peak}}$ represents the forecasted peak occurrence time, and
 \item $T_{\text{observations, peak}}$ represents the observed peak occurrence time.
\end{itemize}

The Peak Timing Error is expressed in discrete time-steps (in this study, each step corresponds to 6-hour intervals). Positive values indicate a delayed forecast (i.e., the model forecasted the peak after the observed occurrence), while negative values reflect a too-early forecast. 
This metric complements magnitude-focused metrics (like RPE and KGE) by highlighting the model’s temporal forecasting capability, allowing identification and potential correction of systematic timing biases in flood forecasting models.

\subsection{Additional metrics for the verification of forecasting models} \label{subsection:WMO metrics}
In accordance with the recently updated Guidelines on the Verification of Hydrological Forecasts~\citep{WMO_2025_Verification_Hydro_Forecasts} issued by the World Meteorological Organization (WMO), we expanded our evaluation framework to incorporate targeted metrics. The following table (Table~\ref{tab:WMO Metrics}) summarizes the most relevant performance metrics, with the first section presenting continuous error metrics and the second section presenting binary verification metrics. For further details please refer to the above cited WMO guidelines.

\begin{table*}[h!]
\centering
\caption{Table reporting the continuous error metrics on the left, and the binary verification metrics on the right} \label{tab:WMO Metrics}
\small
\setlength{\tabcolsep}{3pt}
\begin{tabular}{ll}
\hline
\textbf{Continuous Error Metrics} & \textbf{Binary Verification Metrics} \\
\hline
Mean Absolute Error (MAE) & Probability of Detection (POD) \\
Mean Squared Error (MSE) & Success Ratio (SR) \\
Root Mean Squared Error (RMSE) & Frequency Bias (FB) \\
Mean Error (ME) & Fraction Correct (FC) \\
Pearson Correlation Coefficient ($r$) & Critical Success Index (CSI) \\
Error Variance ($\text{Var}(e)$) & Equitable Threat Score (ETS) \\
Error Standard Deviation ($\text{SD}(e)$) & Peirce’s Skill Score (PSS) \\
Mean Bias Squared (MSE\_mean\_bias2) & False Alarm Ratio (FAR) \\
Second-order Bias (MSE\_second\_order\_bias) & Probability of False Detection (POFD) \\
\hline
\end{tabular}
\end{table*}

To complete the analysis, time-step based $2x2$ contingency tables have also been computed, reporting the number of hits (H), false alarms (FA), misses (M), and true negatives (TN), as defined in the WMO guidelines.

\subsection{Power usage estimation} \label{subsection:method-power-estimate}

To provide a more comprehensive assessment of the framework and its potential implementation and impact, the energy required for delivering forecasts was also estimated. The framework was evaluated on two different hardware configurations: (i) a low-specification workstation equipped with an Intel i7-8700 CPU, 32~GB RAM, and an 80+~Gold certified 300~W power supply unit (PSU), and (ii) a mid-range workstation equipped with an Intel i9-14900K CPU, 192~GB RAM, and an 80+~Gold certified 850~W PSU. Energy consumption was estimated using a workload-weighted power averaging method. Wall-draw power at high and low load was determined from processor utilisation and PSU efficiency benchmarks, corresponding to $\sim$130~W and 60~W for the i7 system, and $\sim$220~W and 100~W for the i9 system. Task durations were derived by experimental observations of different combinations of coding choices, watershed size, and optimization, with mean power calculated as the weighted average across training/forecasting and data I/O phases. Per-task (i.e. per-watershed) energy was then derived by multiplying mean power by task duration, and cumulative energy requirements were obtained by linear extrapolation to the total number of tasks.

\section{Results} \label{section:results}

In the following section a selection of the most relevant results is presented, starting from the hydrological and peak metrics, to the WMO guidelines compliant metrics.

\begin{figure}[h!]
 \centering
 \includegraphics[width=1\linewidth]{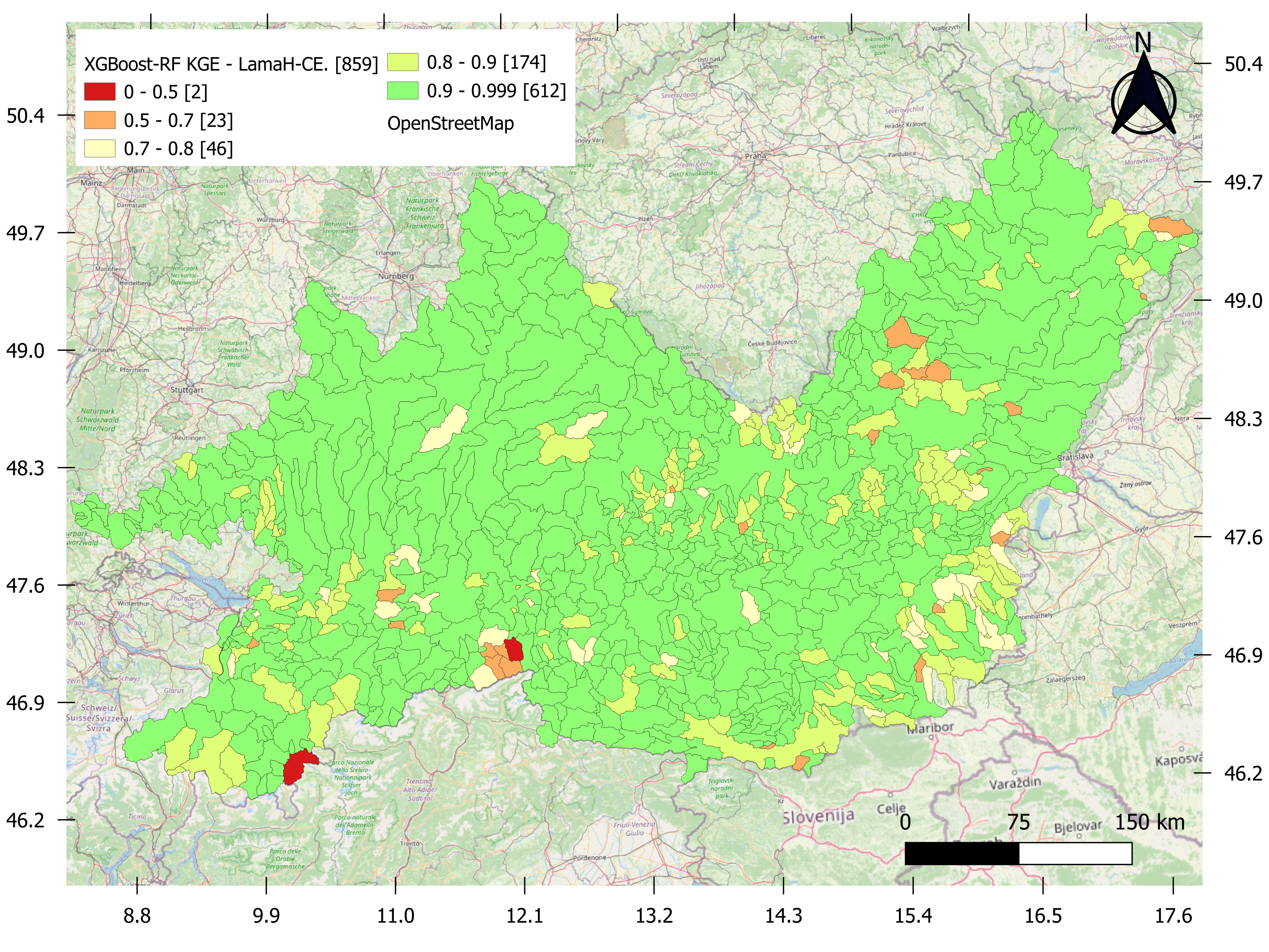}
 \caption{Spatial distribution of the Kling-Gupta Efficiency (KGE) obtained with the XGBoost-RF river-flow forecasting model for 857 LamaH-CE catchments during the testing period. 
 Colour coding: \textbf{red}: KGE < 0.5 (poor); \textbf{yellow}: 0.7 $\leq$ KGE < 0.8 (good); \textbf{light-green}: 0.8 $\leq$ KGE < 0.9 (very good); \textbf{green}: KGE $\ge$ 0.9 (excellent) performance. Map background: OpenStreetMap}
 \label{fig:LamaH-CE_KGE}
\end{figure}

When applied to 857 catchments from the LamaH-CE dataset, the developed XGBoost-RF river discharge forecasting framework produced consistently high KGE values, with only a few localized exceptions. A visual representation of the performance is shown in figure~\ref{fig:LamaH-CE_KGE}. 

In detail, the results indicate a robust transferability skill: 71\% (612 basins) achieve KGE > 0.90, which is regarded as an excellent forecasting performance (light-green in the figure). An additional 20\% (174 basins) have a KGE between 0.80 and 0.90, which is still considered very good (the intermediate light-green-to-yellow shade shown in the figure). A further 5.4\% (46 basins) exhibit KGEs of 0.70 - 0.80 and 2.7\% (23 basins) of 0.60 - 0.70, suggesting some limitations but overall satisfactory skill. Poor performance (KGE < 0.50) is confined to just two alpine basins.

\begin{figure*}
 \centering
 \includegraphics[width=1\linewidth]{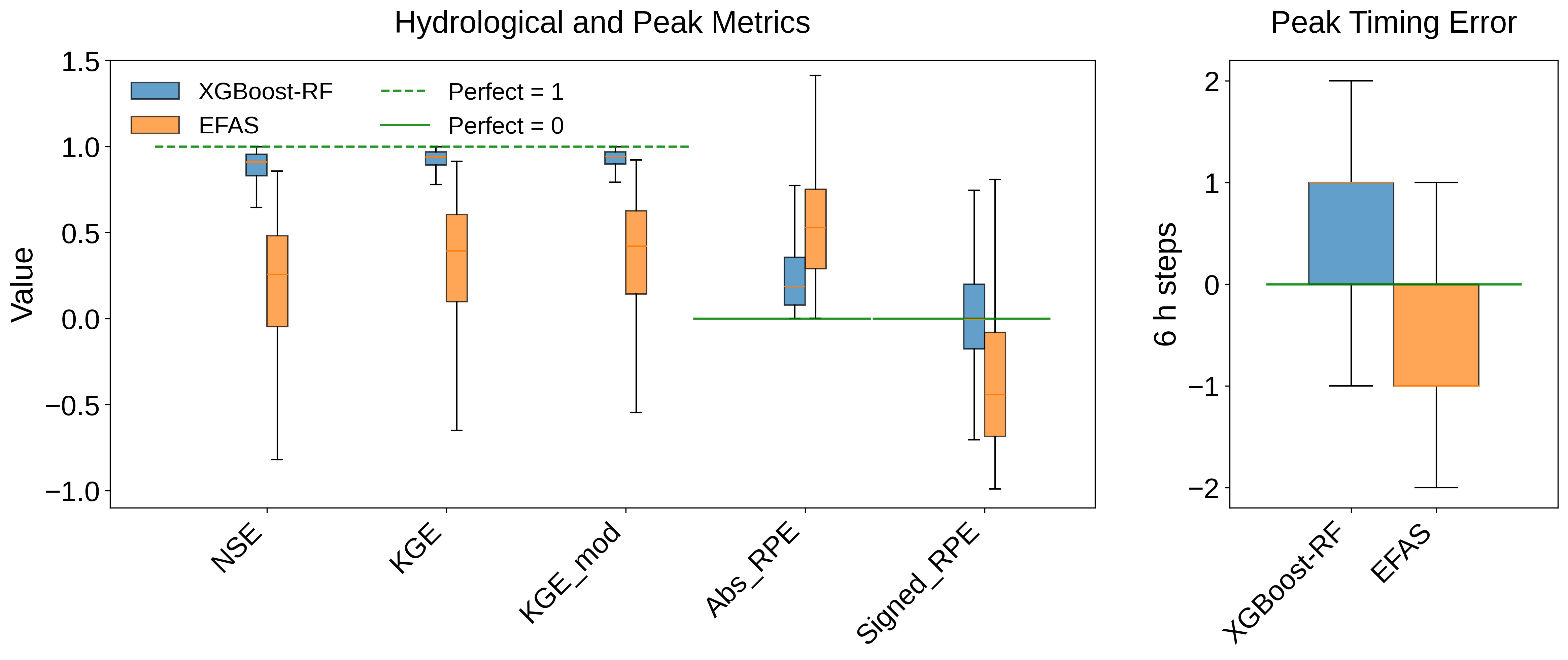}
 \caption{Forecasting performance of the XGBoost-RF model (light blue) versus the EFAS benchmark (orange) across all LamaH-CE catchments. The left panel shows box plots of NSE, KGE, KGE\_mod, Abs\_RPE, and Signed\_RPE; the right panel displays peak-timing errors (6-h steps). XGBoost-RF consistently outperforms EFAS in magnitude metrics and exhibits lower absolute peak-error variability, while its peak-timing error is slightly delayed compared with EFAS.}
 \label{fig:Metrics_panel}
\end{figure*}

We reported the most relevant overall hydrological and peak metrics in figure~\ref{fig:Metrics_panel}, both for XGBoost-RF framework and the EFAS simulations. A dashed green line highlights the maximum theoretical value for NSE, KGE, and KGE'. Conversely, continuous green lines mark the optimal value - zero - for the relative peak errors and for the peak timing error. All the metrics are indicative of good performances with very high KGE and NSE values. The absolute relative peak flow errors (Abs\_RPE) are lower for XGBoost-RF, indicating improved accuracy in peak magnitude forecasts compared to EFAS. The signed relative peak flow errors (Signed\_RPE) reveal that while both models exhibit variability, XGBoost-RF tends to have fewer large errors, thereby reducing large over- or under-estimation of peaks. However, the analysis of Peak Timing Error suggests a slightly different performance pattern: XGBoost-RF tends towards delayed peak forecasts, whereas EFAS forecasts are often slightly early. Overall, these results illustrate the enhanced capability of the XGBoost-RF model for flood forecasting, providing improved forecasting accuracy. 

\begin{figure*}[h!]
  \centering
  \includegraphics[width=1\linewidth]{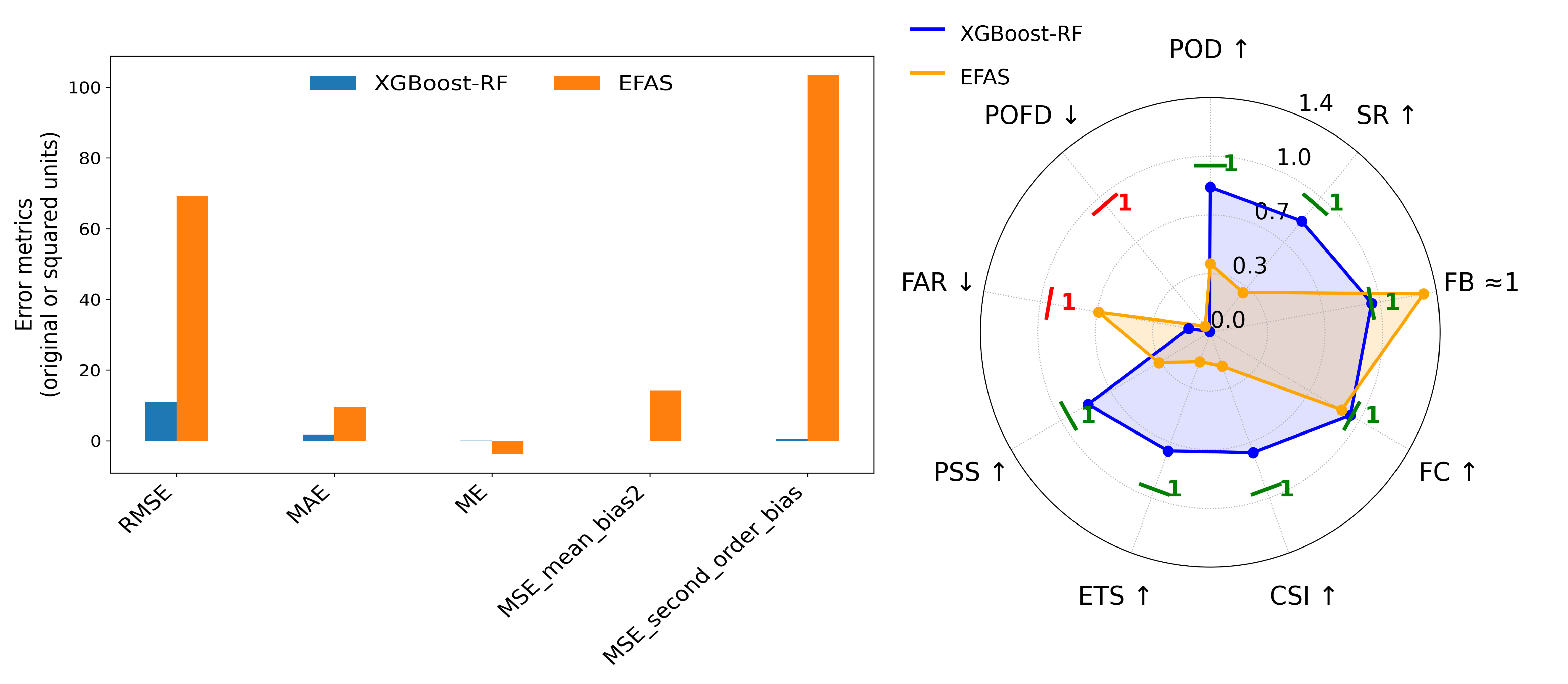}
  \caption{Comparative analysis of hydrological forecasting performance between the XGBoost-RF model and the EFAS benchmark. The left panel displays continuous error metrics (RMSE, MAE, ME, MSE Mean Bias Squared, and MSE Second-order Bias), while the right panel features binary verification metrics (POD, POFD, SR, FAR, FB, FC, CSI, ETS, and PSS) in a radar plot.}
  \label{fig:overall skills}
\end{figure*}

In the left panel of Figure~\ref{fig:overall skills}, we present a selection of the most relevant continuous error metrics as outlined in section~\ref{subsection:WMO metrics}. The XGBoost-RF framework demonstrates a Root Mean Square Error (RMSE) of 10.89 [m\textsuperscript{3}/s], a Mean Absolute Error (MAE) of 1.78 [m\textsuperscript{3}/s], and a Mean Error (ME) of 0.08 [m\textsuperscript{3}/s], all of which exhibit lower values compared to the EFAS benchmark. Additionally, the MSE Mean Bias Squared and the MSE Second-order Bias metrics indicate that XGBoost-RF is nearly bias-free, with minimal deviations of 0.007 and 0.52, respectively.

The right panel of figure~\ref{fig:overall skills} presents binary verification metrics, as defined in section~\ref{subsection:WMO metrics}. The Probability Of Detection (POD) demonstrates a substantial advantage for the XGBoost-RF framework (0.87) over EFAS (0.41), indicating that the former captures more than twice as many observed events. The Probability Of False Detection (POFD) remains low in both models, with XGBoost-RF (0.007) exhibiting significantly lower values than EFAS (0.05). The Success Ratio (SR) further confirms this behaviour, with our framework achieving 0.87 compared to EFAS’s 0.31, showing the latter’s tendency to generate false alarms. The complimentary False Alarm Ratio (FAR) is, as expected, showing EFAS (0.69) to be markedly higher than XGBoost-RF (0.13). In terms of Frequency Bias (FB), EFAS exhibits over forecasting tendency (1.32), whereas XGBoost-RF remains nearly unbiased ($\sim$1). While both models achieve high overall Fraction Correct (FC) scores (0.99 for XGBoost-RF vs. 0.92 for EFAS), skill-based metrics provide a pronounced contrast. The Critical Success Index (CSI) reaches 0.76 for XGBoost-RF while 0.21 for EFAS. Similarly, the Equitable Threat Score (ETS) is 0.75 for the framework and 0.19 for EFAS and finally, the Peirce Skill Score (PSS) further highlights this gap, with XGBoost-RF achieving 0.86 compared to EFAS’s 0.36.

Table~\ref{tab:XGBoost-RF CM} and Table~\ref{tab:EFAS CM} are the $2x2$ contingency tables, which summarize the aggregated results across the entire study area, highlighting the strong forecasting performances of our framework, accordingly with the other metrics reported so far.  

In particular, the XGBoost-RF framework demonstrated superior accuracy, achieving 399497 correctly predicted peak flow events (hits) out of 461651 observed time steps, compared to 189175 hits for EFAS. Conversely, EFAS recorded significantly higher misses (272476) than XGBoost-RF (62154), indicating a marked underestimation of peak flow events. False alarms were also substantially lower for XGBoost-RF (61716) than for EFAS (419965), while true negatives were higher for XGBoost-RF (8692729) than for EFAS (8334480).

\begin{table}
\small
\caption{XGBoost-RF contingency table} 
\centering
\setlength{\tabcolsep}{6pt}
\renewcommand\arraystretch{0.95}
\begin{tabular}{|c|c|c|c|c|}
\hline
\multicolumn{5}{|c|}{\textbf{XGBoost-RF}} \\
\hline
\multicolumn{2}{|c|}{\multirow{2}{*}{\shortstack{2 $\times$ 2\\contingency table}}}
 & \multicolumn{2}{c|}{\cellcolor{cyan!20}\textbf{Observed}} 
 & \multirow{2}{*}{\textbf{Total}} \\
\cline{3-4}
\multicolumn{2}{|c|}{} & \cellcolor{cyan!20}\textbf{Yes} & \cellcolor{cyan!20}\textbf{No} & \\
\hline

\multicolumn{1}{|>{\columncolor{orange!30}}c|}{\textbf{Forecast}}
 & \cellcolor{orange!30}\textbf{Yes}
 & \makecell[c]{H (hits)\\399497}
 & \makecell[c]{FA (false alarms)\\61716}
 & \cellcolor{orange!30}461213 \\
\cline{2-5}
\multicolumn{1}{|>{\columncolor{orange!30}}c|}{}
 & \cellcolor{orange!30}\textbf{No}
 & \makecell[c]{M (misses)\\62154}
 & \makecell[c]{TN (true negatives)\\8692729}
 & \cellcolor{orange!30}8754883 \\
\hline

\multicolumn{2}{|c|}{\makecell[t]{\textbf{Total}}}
 & \cellcolor{cyan!20}461651
 & \cellcolor{cyan!20}8754445
 & \makecell[c]{9216096} \\
\hline
\end{tabular}

\label{tab:XGBoost-RF CM}
\end{table}

\begin{table}
\small
\caption{EFAS contingency table}
\centering
\setlength{\tabcolsep}{6pt}
\renewcommand\arraystretch{0.95}
\begin{tabular}{|c|c|c|c|c|}
\hline
\multicolumn{5}{|c|}{\textbf{EFAS}} \\
\hline
\multicolumn{2}{|c|}{\multirow{2}{*}{\shortstack{2 $\times$ 2\\contingency table}}}
 & \multicolumn{2}{c|}{\cellcolor{cyan!20}\textbf{Observed}}
 & \multirow{2}{*}{\textbf{Total}} \\
\cline{3-4}
\multicolumn{2}{|c|}{} & \cellcolor{cyan!20}\textbf{Yes} & \cellcolor{cyan!20}\textbf{No} & \\
\hline

\multicolumn{1}{|>{\columncolor{orange!30}}c|}{\textbf{Forecast}}
 & \cellcolor{orange!30}\textbf{Yes}
 & \makecell[c]{H (hits)\\189175}
 & \makecell[c]{FA (false alarms)\\419965}
 & \cellcolor{orange!30}609140 \\
\cline{2-5}
\multicolumn{1}{|>{\columncolor{orange!30}}c|}{}
 & \cellcolor{orange!30}\textbf{No}
 & \makecell[c]{M (misses)\\272476}
 & \makecell[c]{TN (true negatives)\\8334480}
 & \cellcolor{orange!30}8606956 \\
\hline

\multicolumn{2}{|c|}{\makecell[t]{\textbf{Total}}}
 & \cellcolor{cyan!20}461651
 & \cellcolor{cyan!20}8754445
 & \makecell[c]{9216096} \\
\hline
\end{tabular}
\label{tab:EFAS CM}
\end{table}

\subsection{Example watershed: ID 332 - Traisen river at Windpassing, Austria}

\begin{figure*}[h!]
  \centering
  \includegraphics[width=1\linewidth]{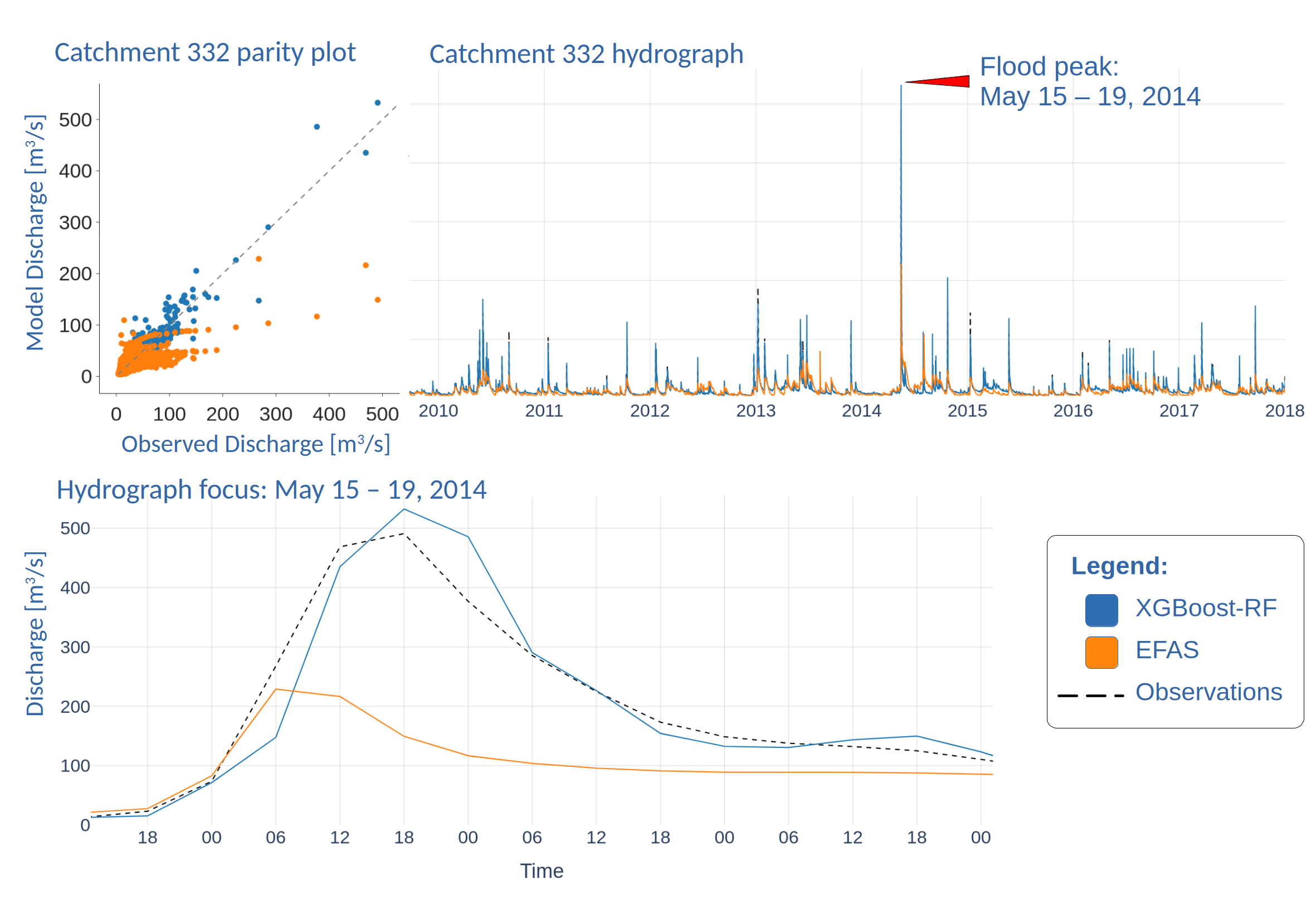}
  \caption{XGBoost-RF and EFAS forecasts for LamaH-CE watershed ID 332 in the period from 2010 to 2018. Left panel: Parity plot of the modelled flow against the observed flow. Right panel: flood hydrograph of the forecasted / observed time series, with a red flag highlighting the rarity of the analysed peak flow event. The lower panel shows the hydrograph of the record peak flow at the Windpassing river gauge (Austria), which occurred on 16 May 2014. For all the panels, XGBoost-RF framework results are reported in blue, while EFAS modelled values are reported in orange. Observations are displayed with a dashed line in the hydrograph plot.}
  \label{fig:332 focus}
\end{figure*}

In addition to the overall metrics evaluation, we randomly selected some watersheds to examine the forecasting performance more in detail, and we report here the findings for the example watershed closed by the gauge with ID = 332 in the LamaH-CE dataset, at Windpassing, Austria. In the top left panel of figure~\ref{fig:332 focus}, is displayed a parity plot comparing both the XGBoost-RF (blue markers) and EFAS (orange markers) forecasts with the observed river discharge. 

XGBoost-RF is able to represent the highest peak flow event of the testing time series, although with a tendency to slightly over-predict. In the right panel we reported the testing time series, highlighting the flood peak of the 16 of May 2014, for which we provided a more detailed hydrograph in the bottom part.
For each 6-hours forecasting time step, the forecasted values are aligned to the observed ones, meaning that at time $t$, the observed discharge and the forecasted values generated one step before at ($t-1$) for that time step ($t$) are synchronously displayed. Even if slightly lower discharge values are forecasted before the peak occurs, the XGBoost-RF captures the flood event very well. For example, when an observer assesses the situation at 06:00, three values are considered: the observed discharge at 06:00; the discharge forecast issued at 00:00 for 06:00 (which, in this example is lower than the observed value, though still indicative of a rising trend); and the forecast issued at 06:00 for 12:00, which accurately captures the steep increase expected over the subsequent six hours. This information is crucial for the authorities to make informed decisions for the emergency management. Moreover, the forecasted value for time 12 was 435 m\textsuperscript{3}/s, with respect to the observed value of 468 m\textsuperscript{3}/s, demonstrating a very small error, less than 1\%, on very high and unusual flows for that time series. 

\subsection{Energy consumption estimation} \label{subsection:energy}

The execution time of a single training-forecasting task (1 watershed) ranged between 30~seconds and 4~minutes, depending on watershed size, code implementation, and optimisations. 
Corresponding energy use estimation varied across the two tested hardware configurations, and is presented in table~\ref{tab:energy-estimate}. 
On the i7 - 8700 workstation, average consumption per task was estimated between 0.0007 and 0.0056~kWh, leading to a cumulative requirement of 0.6 - 4.9~kWh for 857 tasks. 
On the more powerful i9 - 14900K system, per-task consumption increased to 0.0012--0.0093~kWh, corresponding to 1.0 - 8.0~kWh for 857 tasks. 
Although the energy use per task remains modest (well below 0.01~kWh), scaling to hundreds of watersheds results in substantial cumulative requirements, particularly for the higher-specification system, where the lack of significant time improvement due to I/O bottlenecks leads to proportionally higher energy demand. 

\begin{table}[h!]
\centering
\small
\setlength{\tabcolsep}{1.5pt}
\caption{Estimated energy requirements for delivering forecasts, per task and for 857 tasks, under two hardware configurations. Values are given as ranges corresponding to task durations between 30~s and 4~min.}
\begin{tabular}{@{}lcc@{}}
\textbf{Workstation} & \textbf{Per task (kWh)} & \textbf{857 tasks (kWh)} \\
\hline
i7--8700 (32~GB, 300~W PSU) & 0.0007 -- 0.0056 & 0.6 -- 4.9 \\
i9--14900K (192~GB, 850~W PSU) & 0.0012 -- 0.0093 & 1.0 -- 8.0 \\ 
\hline
\end{tabular}
\label{tab:energy-estimate}
\end{table}

\section{Discussion} \label{section:discussion}
 
Rigorous modelling of rainfall-runoff processes requires access to high-quality, long-term historical data for many hydrological, geomorphological, and meteorological variables. Retrieving this data is challenging, and processing it within operational forecasting time frames requires substantial computational resources, typically available only on High-Performance Computing (HPC) systems like those used by the EFAS system for running the LISFLOOD model to generate flood forecasts~\citep{ecmwf_fact_sheet_2022, ecmwf_hydrology_2021, ecmwf_supercomputer_facility}.
We addressed data-related issues by limiting the input variables for our framework to rainfall and river discharge, which are the primary cause and effect in the rainfall-runoff process for the majority of the watersheds analysed. Moreover, we enabled XGBoost and RF to learn the unique behaviour of each single catchment by training them for the catchment-specific observed time series. This enables the models to implicitly account for hidden hydrological processes related to variables not explicitly considered but affecting runoff generation. This also allowed us to avoid issues related to considering static layers (e.g., geology, elevation) in ML methods (as introduced in section~\ref{section:introduction}). Additionally, transferring the models' hyperparameters, architecture, and calibration across the 857 catchments allowed us to limit the computational resources required to cover the 170000 km\textsuperscript{2} of the LamaH-CE domain, making it feasible on workstation hardware. This is also achieved thanks to the architecture of the XGBoost and Random Forest (RF) methods, which are inherently efficient and powerful even when running on CPU. Although both XGBoost and, more indirectly, RF, can be trained on GPUs, in our case, the potential speed-up is limited by the I/O operations required to read and write the time series for each watershed, which represent one of the major constraints on computational performance, even in broader contexts. Regarding the rainfall, by assuming a single value for the entire watershed for each time step, we ignored the rainfall internal spatial variability which can affect runoff generation, especially in larger watersheds; however, the model compensates by incorporating multiple time steps of rainfall data, as discussed in Section~\ref{subsubsection:lag-roll}, thus indirectly accounting for this variability.

Regarding the capability to forecast peak events, the combination of a binary flag with the \texttt{peak\_magnitude} variable allows both the detection and quantification of extreme discharge events. This approach is particularly effective not only for training but also for evaluating model performance for rare hydrological conditions. The choice of a quantile-based threshold aligns with standard practices in hydrological extremes' analysis, providing a non-parametric, clear, and reproducible method for isolating and characterizing critical high-flow episodes. Carefully specializing an RF model on these selected events gives our framework the unique capability to focus on and learn the data patterns usually associated with peaks, thus improving the forecasts of rare events.
High KGE values for flood forecasts are achieved in most catchments in the LamaH-CE domain without pronounced regional biases, indicating that the modelling framework is highly adaptable to various regions. Lower-performing areas tend to cluster in regions of high elevation or complex hydrological settings, where glacial-nival dynamics and perennial snow cover are likely to affect the forecasting accuracy of rainfall-driven models. Therefore, additional analyses and process-specific representations are required (e.g., including snow and temperature). 
Moreover, the detailed example of Windpassing catchment and the contingency matrices indicate that the proposed XGBoost-RF framework provides forecasts that are both accurate and operationally reliable, even compared to the EFAS benchmark. These improvements are particularly valuable for operational hydrological forecasting, where timely identification of flow peaks with low false alarm rates is crucial for effective flood risk management and emergency response planning.
Regarding the comparison with EFAS forecasts, it should be noted that EFAS' simulations are not intended for short-term, high-detail local-scale forecasting but are designed for longer forecasting windows at large spatial scales using a completely different approach. XGBoost-RF forecasting framework may complement the EFAS and other systems for short-range forecasts.
Regarding the further evaluation metrics adopted to test the XGBoost-RF framework, it demonstrates a consistently high level of skill across all binary verification metrics, confirming that it captures the essential dynamics of flood occurrence while producing reliable, unbiased forecasts. The high POD (0.87) indicates that the system detects the vast majority of observed flood events. A high Success Ratio (0.87) and a small False-Alarm Ratio (0.13) underline that most alarms issued by the model correspond to real events, a seminal characteristic for maintaining trust among users. The near-neutral Frequency Bias ($\sim$1) further confirms that the model neither over- nor under-predicts the number of floods, while the exceptionally high Peirce Skill Score (0.86) remarks a balanced improvement in both hit and false-alarm performance. In addition, skill-based indices such as CSI (0.76) and ETS (0.75) further underline the framework’s ability to combine strong detection with low false-alarm rates, offering a robust overall assessment of forecast quality.
From an operational standpoint, these results represent tangible benefits for flood warning services. High detection capability ensures that potentially hazardous events are flagged in advance, allowing authorities to activate mitigation measures (e.g., evacuations, reservoir releases). Simultaneously, the low incidence of false alarms reduces unnecessary social and economic disruption, preserving the credibility of the warning system and improving and maintaining trustfulness among stakeholders. The near-unbiased frequency response also simplifies downstream integration with hydrodynamic routing or inundation-mapping tools, as the forecasts can be used directly without extensive post-processing adjustments. 
Power usage estimates highlight that, although the per-task energy use is very small in absolute terms, cumulative requirements scale with the number of watersheds processed. Moreover, the higher-specification i9 system, despite its greater computational capacity, did not yield significantly shorter runtimes due to I/O bottlenecks, resulting instead in proportionally higher energy demand. This indicates that optimising data transfer and storage operations may offer greater efficiency gains than further increases in raw CPU performance. When compared to high-performance computing (HPC) facilities or recent reports on the energy footprint of AI training, the absolute consumption reported here is very low; nevertheless, the same scaling principles apply. As frameworks like this are extended to larger domains or coupled with machine learning components, careful attention to computational efficiency and energy demand becomes critical for sustainable implementation.

\section{Conclusions} \label{section:conclusion}

In this study, we introduced a lightweight framework which combines and leverages two powerful machine learning models: XGBoost for streamflow forecasting, combined with a Random Forest specialized in peak-flow forecasting. The framework was set up for six-hour streamflow forecasts on pilot watersheds, and then transferred to 857 LamaH-CE catchments. Our XGBoost-RF framework is designed to be simple and effective, allowing it to remain locally sensitive but transferable to larger scales, while being trained and applied efficiently to generate forecasts on standard computing hardware.
Across the full LamaH-CE domain, XGBoost-RF consistently outperformed EFAS: errors were smaller, and correct peak detection exceeded 87\%, with a false-alarm rate around 13\%. The peak timing of the forecasted time series tends to shift slightly forward in time, yet all randomly and manually checked examples demonstrated that the forecasted values are anyway marking the flood wave in advance, even if the absolute peak might be slightly shifted from the observed one. Further work is underway to improve alignment.
The results confirm that an ad-hoc engineered data-driven machine learning framework can maintain local accuracy while being easily deployable and transferred to large spatial scales, also complementing large-scale process-based forecasting systems (e.g., EFAS), which have coarser resolution and use simplified process models (e.g., flood routing), inherently limiting their performance in smaller catchments~\citep{thielen09_efas, Alfieri2013_GloFAS}.
Compared with recent deep-learning models that rely on sequence-to-sequence architectures, graph neural networks, or attention mechanisms~\citep{kratzert19_flood, chen23_gnn}, the XGBoost and the RF architecture proposed in this study are extremely lightweight, and train in seconds on a standard desktop computer CPU, also avoiding the computationally and energy expensive hyper-parameter searches typically required by more complex data-based modelling approaches.
Several limitations persist. While the 6-hour forecasting horizon currently offers valuable and timely insights, particularly in facilitating the deployment of emergency measures, extending this time frame while maintaining high levels of forecast accuracy and reliability, could yield greater benefits. Such extension has the potential to transform unforeseen emergencies into anticipated and more manageable situations. Future work will focus on developing an enhanced framework that incorporates additional variables and extends the forecasting horizon to improve predictive capabilities. The current framework is designed to work in data-rich areas to gain high-quality, sound knowledge of a large variety of watersheds. However, data availability is not evenly distributed, and there are many areas in which valuable historical data and continuous monitoring, as provided by LamaH-CE and EMO-1, cannot be leveraged. This is the subject of future research. Moreover, a finer calibration of the lag features and rolling statistic might be useful to better specialize the model for different watersheds. 

Despite these considerations, the XGBoost-RF framework provides a reliable, high-skill foundation for future flood forecasting. The performance obtained by the proposed XGBoost-RF short-term flood forecasting framework underlines that blending data science and hydrological knowledge allowed for reaching enhanced forecasting outcomes. The hydrological conceptualization of the problem and the machine learning solution developed improved forecast accuracy and reliability for numerous watersheds, delivering valuable information for real-world operational flood forecasting. 

\section*{Acknowledgements}

This research has been supported by the Italian National Recovery and Resilience Plan (PNRR), European Union - NextGenerationEU, Ministerial Decree No. 351/2022.

\section*{Declaration of interest}

The authors declare that they have no known competing financial interests or personal relationships that could have appeared to influence the work reported in this paper.

\section*{Author contributions}
\noindent
Gabriele Bertoli: Conceptualization, Data curation, Formal analysis, Investigation, Methodology, Writing – original draft, Writing – review \& editing.\\  
Kai Schröter: Conceptualization, Supervision, Writing – review \& editing.  \\
Rossella Arcucci: Conceptualization, Supervision, Writing – review \& editing.  \\
Enrica Caporali: Conceptualization, Supervision, Writing – review \& editing.\\


\bibliography{main}

\end{document}